\documentclass[prl,twocolumn,floatfix, showpacs]{revtex4}%

\usepackage{amsfonts}
\usepackage{amsmath}
\usepackage{amssymb}
\usepackage{graphicx}%

\begin{document}

\preprint{cond-mat/000000}

\title{Nodal Order Parameter in Electron-Doped Superconducting Films\\ $\mathrm{Pr_{2-x}Ce_xCuO_{4-\delta}}$ ($x=0.13$,
$0.15$ and $0.17$)}

\author{A.~Snezhko}
\author{R.~Prozorov}
\affiliation{Department of Physics \& Astronomy and USC NanoCenter, \\ University of South Carolina, 712 Main
St, Columbia, SC 29208.}

\author{D.~D.~Lawrie}
\author{R.~W.~Giannetta}
\affiliation{Loomis Laboratory of Physics, University of Illinois at Urbana - Champaign, \\1110 West Green
Street, Urbana, Illinois 61801.}

\author{J.~Gauthier}
\author{J.~Renaud}
\author{P.~Fournier}
\affiliation{Canadian Institute of Advanced Research, \\Centre de recherche sur les propri\'et\'es
\'electroniques de mat\'eriaux avanc\'es, D\'epartement de Physique, Universit\'e de Sherbrooke, Sherbrooke,
Qu\'ebec, CANADA, J1K 2R1.}

\keywords{type-II superconductivity, penetration depth, irreversibility}

\pacs{74.25.Nf,74.72.Jt}

\begin{abstract}

The London penetration depth, $\lambda _{ab}$(T), is reported for
thin films of the electron-doped superconductor
Pr$_{2-x}$Ce$_{x}$CuO$_{4-\delta }{\rm }$ at three doping levels
(x = 0.13, 0.15 and 0.17). Measurements down to 0.35 K were
carried out using a tunnel diode oscillator with excitation fields
applied both perpendicular and parallel to the conducting planes.
For all samples and both field orientations $\lambda _{ab}$(T)
showed power law behavior implying a superconducting gap with
nodes.
\end{abstract}


\received[Received: ]{10 April, 2003}%

\maketitle

It is now generally accepted that the hole-doped high-Tc cuprates
exhibit d-wave pairing symmetry \cite{harlingen,tsuei}. For
electron doped materials, several different experiments have
recently provided strong evidence for unconventional pairing.
These include penetration depth on single crystals
\cite{prozorov,kokales}, tricrystal measurements \cite{tsuei2},
ARPES \cite{armitage}, tunneling \cite{hayashi, biswas}, specific
heat \cite{balci} and Raman scattering \cite{blumberg}.   However,
the situation in thin films has been controversial. Tunnelling
measurements provided evidence for an energy gap \cite{alff}.
Mutual inductance measurements of the penetration depth in
Pr$_{2-x}$Ce$_{x}$CuO$_{4-\delta }{\rm }$ (PCCO) films claimed
gapped behavior in optimal and overdoped films and a transition to
a $T^2$ power law for underdoped films, indicative of d-wave
pairing \cite{skinta}. More recent measurements on PCCO films
grown with a $\mathrm{Pr_2CuO_4}$ buffer layer indicate a nodeless
order parameter at all doping levels \cite{kim}, albeit with an
unusually small value of the superconducting gap ($\sim 0.9T_{c}$
compared to $1.76~T_c$ of a weak-coupling BCS value).

To help resolve this controversy we have measured the penetration
depth in PCCO films for several different doping levels and for ac
magnetic field excitation both parallel and perpendicular to the
film plane. This approach helps to eliminate complications that
might arise from vortex motion and demagnetization effects.

Thin PCCO films of thickness 2500 {\AA} were grown using pulsed
laser deposition on LaAlO$_3$ and yttria stabilized zirconia
substrates. Details of the film growth can be found in
Ref.~\cite{maiser}. The films have been optimized for oxygen
content by maximizing $T_c$ for each cerium concentration. The
films were characterized by x-ray, ac susceptibility and
resistivity measurements. Corresponding critical temperatures and
cerium doping levels for underdoped, optimally doped and overdoped
films used in the measurements are $T_c=9.1$~K ($x=0.13$),
$T_c=20.5 $~K ($x=0.15$) and $T_c= 11.8 $~K ($x=0.17$),
respectively. Penetration depth measurements were carried out
using a 13 MHz tunnel-diode LC resonator described previously
\cite{carrington,prozorov2}. Samples were placed on a movable
sapphire stage with temperature control from 0.35 to 40K mounted
in a $^{3}$He refrigerator. When the magnetic field is applied
perpendicular to the conducting Cu-O planes ($H_{ac} \bot ab)$
only in-plane screening currents are induced. In that case, it can
be shown that the frequency shift of the resonator is proportional
to the change of the in-plane penetration depth, $\Delta
f=\,f\left( T \right)-\,f\left( 0 \right)\,=G\left( {\lambda _{ab}
\left( T \right)\,-\,\lambda _{ab} \left( 0 \right)} \right)$
where $G$ is geometrical factor that depends on the sample and
coil geometry [16, 17]. When $H_{ac}
\parallel ab$ there are no demagnetization effects and the oscillator frequency shift is accurately proportional to the susceptiblity, \cite{prozorov2},
\begin{equation}\label{ } -4\pi \chi \,=\,1\,-\,\,\left(
{{2\lambda _{ab} } \mathord{\left/ {\vphantom {{2\lambda _{ab} }
d}} \right. \kern-\nulldelimiterspace} d} \right)\tanh \left(
{{\,d} \mathord{\left/ {\vphantom {{\,d} {2\lambda _{ab} }}}
\right. \kern-\nulldelimiterspace} {2\lambda _{ab} }} \right)
\end{equation}In principle, $\lambda _{c}$ also contributes to the susceptibility, but its relative contribution is proportional to
\emph{d/w}, where $d$ is a thickness and $w$ is a width
\cite{prozorov2}. Since \emph{d/w}$\sim $10$^{-4}$ this
contribution is negligible.

\begin{figure}[tb]
\includegraphics[width=8.5cm]{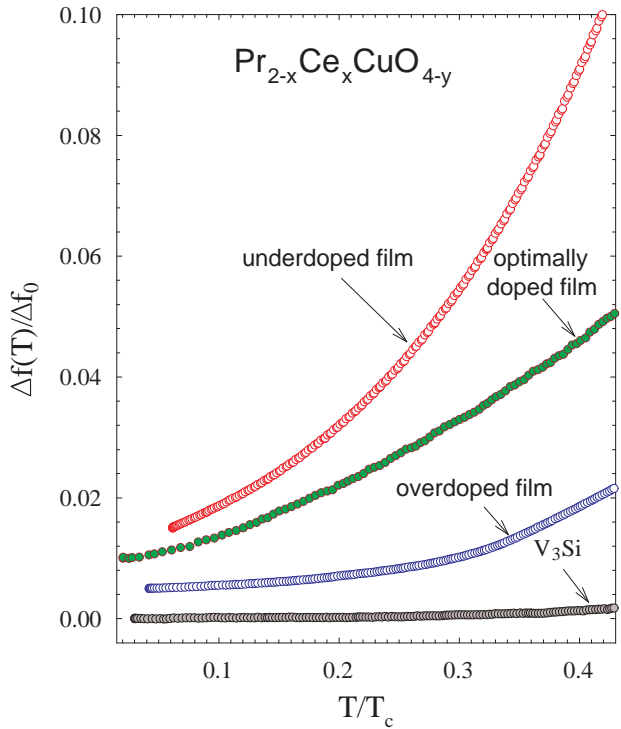}%
\caption{Variation of London penetration depth as a function of
normalized temperature $T/T_c$ for three measured PCCO films of
different Ce doping levels. Data offset was used for clarity. Data
taken for $\mathrm{V_3Si}$ in the same apparatus is an example of
isotropic s-wave behavior. $\Delta f_0 $ is the total frequency
shift when the perfectly diamagnetic sample is
inserted into a resonator at the lowest temperature.}%
\label{fig1}%
\end{figure}
In Fig.~\ref{fig1} we plot the change in penetration depth as a
function of normalized temperature $(T/T_{c})$ for three measured
PCCO films with differing \textit{Ce} doping levels (ac field was
applied perpendicular to the films plane).  Owing to the very
large demagnetization factor, it is difficult to accurately
determine the calibration constant G relating the frequency shift
to the change in penetration depth.  We therefore plot the
normalized frequency shift alone. For comparison we show data
taken for V$_3$Si, a weak-coupling isotropic $s$-wave
superconductor, taken in the same apparatus. Small data offsets
were used for clarity. $\Delta f_0 $ corresponds to the total
frequency shift when the perfectly diamagnetic sample is inserted
into a resonator at the lowest temperature.   The low temperature
behavior of all three measured PCCO samples differs from that
obtained for the s-wave material. However, there are considerable
differences in behavior between PCCO films of different
\textit{Ce} doping levels.

\begin{figure}[b]
\includegraphics[width=8.5cm]{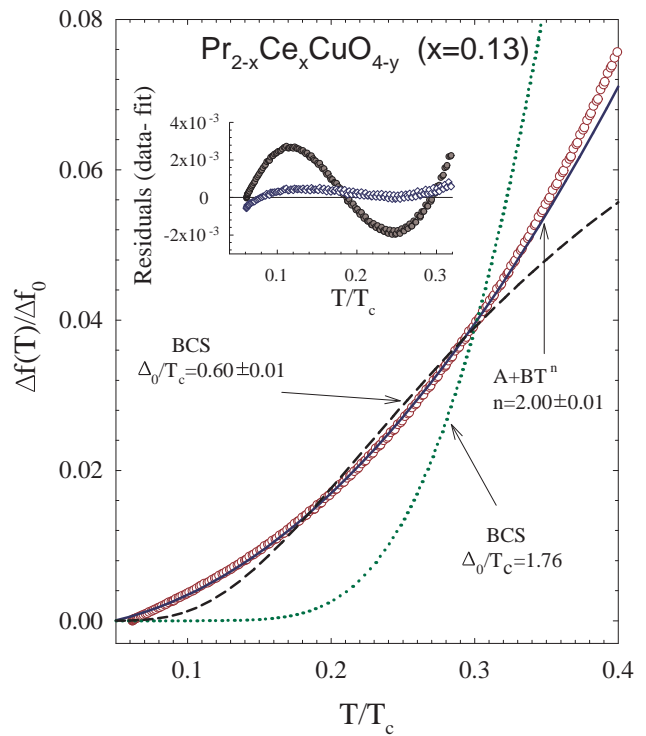}%
\caption{Low temperature penetration depth variation in underdoped
PCCO film. Lines show fits to different models described in the
text. The inset shows residuals (experimental data -- fit) for the
best BCS (solid symbols) and best power law (open symbols) fits.
The low temperature exponential s-wave fit is unacceptably
poor. The best fit is achieved for $T^n$ power law with $n$=2.00$\pm $0.01}%
\label{fig2}%
\end{figure}

The data for an underdoped PCCO film is shown in Fig.~\ref{fig2},
along with several fits. The fitting range was chosen up to
0.32T$_{c }$ to ensure the validity of low temperature BCS
expression for an s-wave material,
\begin{equation}\label{2}
\Delta \lambda =\lambda (0)\sqrt {\frac{\pi \Delta _0 }{2T}} \exp
\left( {\frac{\Delta _0 }{T}} \right)
\end{equation}
Here $\Delta _0 $ is the value of the energy gap at zero
temperature [18]. In each case the small negative offset
$A=\lambda (0)-\lambda (0.4K)$ was determined as a fitting
parameter. The dotted line shows a fit to the standard BCS form
with the weak coupling gap value $\Delta _0 /T_c $ = 1.76.  The
dashed curve shows the best fit obtained by allowing $\Delta _0 $
to be a free parameter.  In this case the extracted value $\Delta
_0 /T_c =0.60\pm 0.01$ is considerably smaller than the
weak-coupling BCS value.  The solid line is a fit  to a power law,
$\Delta f \sim A+BT^{n}$.  The best result was achieved with
$n$=2.00 $\pm $ 0.01. The inset displays residuals for the various
fits. The T$^{2}$ fit is superior to either s-wave fit, suggesting
a gap function with line nodes and unitary limit impurity
scattering, as reported earlier for PCCO crystals [19-21].

\begin{figure}[tb]
\includegraphics[width=8.5cm]{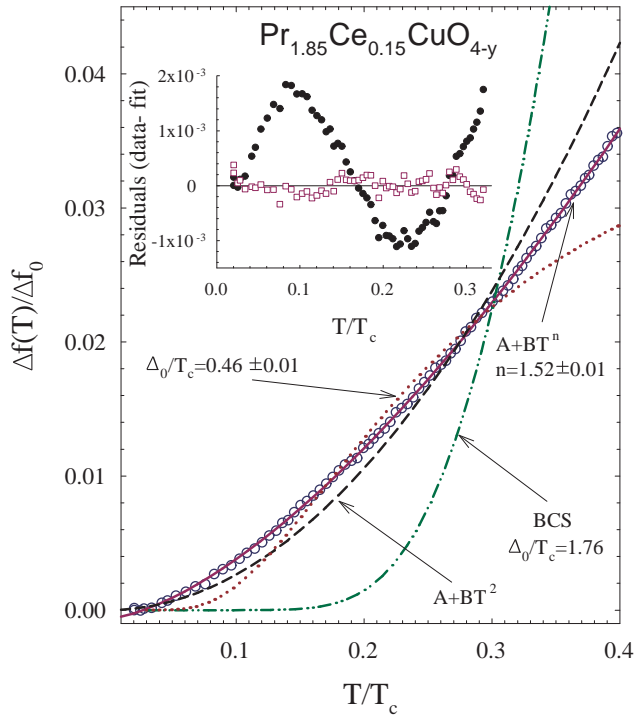}%
\caption{Magnetic penetration depth as a function of normalized
temperature for optimally doped PCCO film. Lines show fits to
different models. The best BCS fit yields low $\Delta _0 /T_c
=0.46\pm 0.01$. The best fit is achieved for $T^{n}$ power law
with $\emph{n}=1.52\pm $0.01. Inset: residuals, experimental data
- fit, for the best BCS (solid circles) and best power law (open
squares) fits.}%
\label{fig3}%
\end{figure}

Fig.~\ref{fig3} shows the same analysis for an optimally doped
PCCO film.  The dash-dotted line in the figure represents the BCS
fit with $\Delta _0 /T_c $ = 1.76 while the dotted line shows the
best fit with $\Delta _0 $ as free parameter.  In this case we
obtain $\Delta _0 /T_c =0.46\pm 0.01$.  The dashed line is a fit
to the power law, $\Delta \lambda \sim A+BT^{2}$. Allowing the
exponent to be a free parameter, the best fit for optimally doped
PCCO film was achieved for $\Delta \lambda \sim T^{1.52\pm 0.01}$
, displayed in the graph as solid line. The residuals plot (inset)
indicates that a power law provides a superior fit to an
exponential. Although certain organic superconductors do
consistently exhibit a $\lambda \sim T^{1.5}$ power law behavior
~\cite{carrington}, in this case the $n$ = 1.5 power is more
likely to be an alternative way to describe the well-known "dirty
d-wave" model in which $ \lambda \sim T^2/(T^{*} + T)$ where
$T^{*}$ is a measure of the rate of unitary limit impurity
scattering [19-21]. A fit to this form with $T^{*} = 0.29 T_{c}$
gives a fit of nearly identical quality to the fractional power
law.
\begin{figure}[tb]
\centerline{\includegraphics[width=8.5cm]{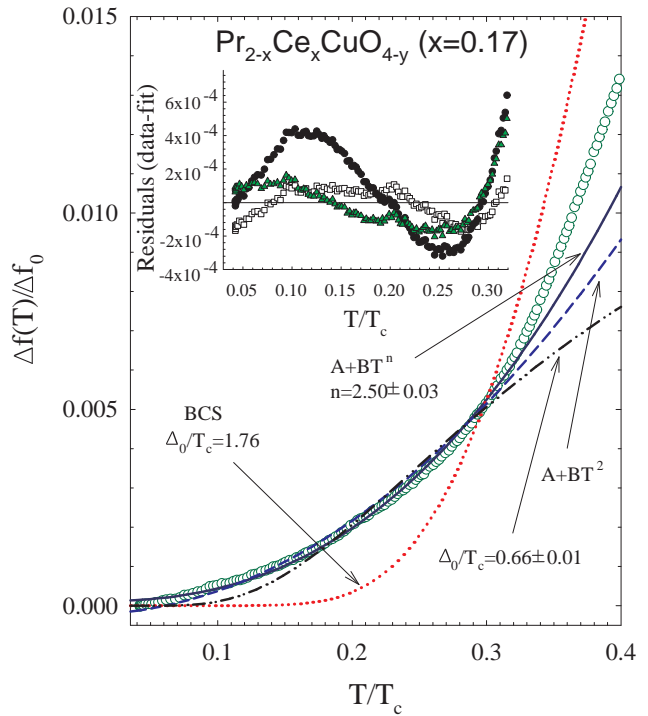}}%
\caption{Penetration depth behavior in overdoped PCCO film. Lines
show fits to different models described in the text. The best fit
is achieved for $T^n$ power law with $n$=2.50$\pm $0.03. Inset
represents residuals, experimental data -- fit, for the best BCS
fit (solid circles)
and power law fits: with n=2.50 (open squares) and n=2 (solid triangles).}%
\label{fig4}%
\end{figure}
\begin{figure}[tb]
\includegraphics[width=8.5cm]{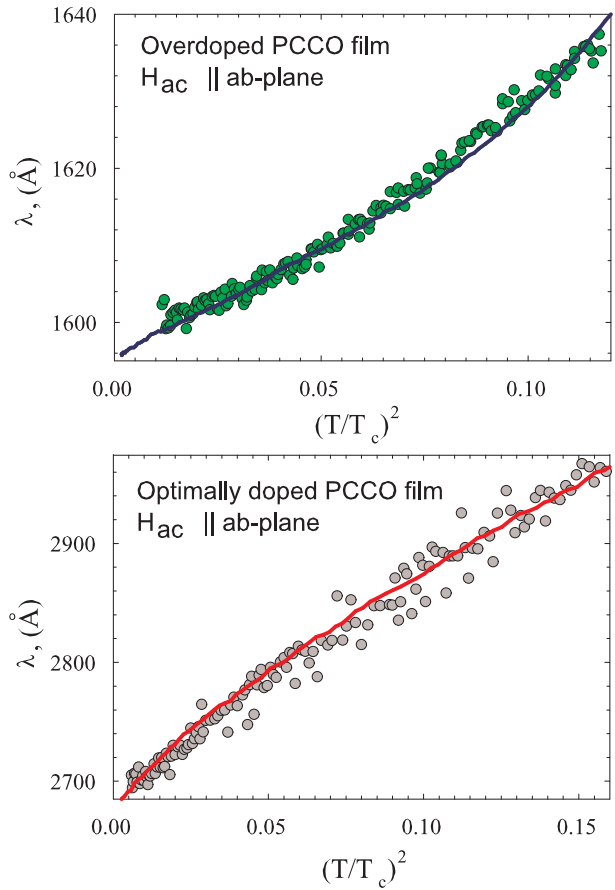}%
\caption{Penetration depth as a function of squared normalized
temperature for overdoped (top graph) and optimally doped (bottom
graph) PCCO films measured in parallel orientation with respect to
Cu-O planes
($H_{ac}\parallel a)$. Solid lines represent results obtained for H$_{ac}$ oriented perpendicular to the films plane.}%
\label{fig5}%
\end{figure}

Fig.~\ref{fig4} shows the data for an overdoped PCCO film. With
$\Delta _0 $ as a free parameter we obtain $\Delta _0 /T_c
=0.69\pm 0.01$. The dotted line shows the fit for $\Delta _0 /T_c
=1.76$. The dashed line shows the fit to a "dirty d-wave"
expression, $\Delta \lambda \sim A+BT^2$. Finally, the solid line
shows the fit to the power low expression, $\Delta \lambda \sim
A+BT^n$. The best overall fit was achieved with for for $n$=2.50
$\pm $ 0.03.

In the figure \ref{fig5} we plot the change in penetration depth
as a function of $(T/T_{c})^{2}$ obtained for optimally and
overdoped PCCO films measured in the parallel configuration
($H_{ac} \parallel a)$.  In this case, the absence of
demagnetizing effects allows us to determine the change in
$\lambda$ directly from the frequency shift.  Since $\lambda$ is
comparable to the film thickness the total frequency shift in this
geometry is much less than in the previous orientation and the
data is correspondingly noisier.   The solid lines through the
data are not fits but represent the scaled data taken in the
($H_{ac} \perp a)$  orientation with an additive offset for the
$\lambda_0$. The overall temperature variation is the same in both
orientations, giving us confidence that only the in-plane
penetration depth is being measured.

To summarize, we find that London penetration depth in underdoped,
optimally doped and overdoped PCCO films studied can be fit by a
dirty d-wave temperature dependence rather than a gapped order
parameter. Indeed, it is possible to observe apparently
non-exponential behavior of $\lambda(T)$ for samples with
extremely large spread of transition temperatures due to
inhomogeneities in chemical composition. However, our numerical
solution in the framework of the weak-coupling s-wave BCS theory
indicates that in order to mimic the $T^2$ behavior observed, the
sample would have to contain a linear probability distribution of
$T_c$s extending from $T_c(max)$ down to 0K.  Such a distribution
is chemically unfeasible and there is no indication of chemical
inhomogeneity from the observed superconducting transition widths.
Furthermore, good agreement of the thin film data with the earlier
data obtained on single crystals indicates the unlikeness of this
scenario.   The $n = 2.5$ effective  power law observed in
overdoped films is difficult to explain with any simple model of
the gap function.   However, it is clear that the disparity
between the power law and the finite gap model is smaller for
overdoped films and could indicate a progression towards the
formation of a small gap with \emph{Ce} doping or for a particular
kind of substrate.

Acknowledgments: the work at University of South Carolina was
supported by the NSF/EPSCoR under Grant No. EPS-0296165. Work at
the University of Illinois was supported by NSF under grant No.
DMR-0101872. PF acknowledges the support of Canada Foundation for
Innovation, the Natural Sciences and Engineering Research Council
of Canada, and the Foundation of the Université de Sherbrooke.


\begin{thebibliography}{99}

\bibitem{harlingen}D. Van Harlingen, Rev. Mod. Phys. 67, 515 (1995)

\bibitem{tsuei}C.~C. Tsuei, J.R. Kirtley, Rev.Mod.Phys. 72, 969 (2000)

\bibitem{hayashi}F.~Hayashi \textit{et al}., J. Phys. Soc. Jpn. 67, 3234 (1998)

\bibitem{prozorov}R.~Prozorov, R.W. Giannetta, P. Fournier, R.L. Greene, Phys. Rev. Lett.
85, 3700 (2000)

\bibitem{kokales}J.~D. Kokales \textit{et al.}, Phys. Rev. Lett. 85, 3696 (2000)

\bibitem{tsuei2}C.~C.Tsuei, J.R. Kirtley, Phys. Rev. Lett. 85, 182 (2000)

\bibitem{armitage}N.~P. Armitage \textit{et al}., Phys. Rev. Lett. 85, 3696 (2000)

\bibitem{biswas}A. Biswas \textit{et al}., Phys. Rev. Lett. 88,
207004 (2002)

\bibitem{balci}H. Balci \textit{et al}., Phys. Rev. B 66, 174510
(2002)

\bibitem{blumberg}G.~Blumberg \textit{et al}., Phys. Rev. Lett. 88 107002 (2002)

\bibitem{alff}L.~Alff \textit{et al.}, Phys. Rev. Lett. 83, 2644 (1999)

\bibitem{skinta}J.~A Skinta \textit{et al}., Phys. Rev. Lett. 88, 207005 (2002)

\bibitem{kim}Mun-Seog Kim, J.~A. Skinta, and T.~R. Lemberger, cond-mat/0302086
(2003)

\bibitem{maiser}E. Maiser et al., Physics 297C, 15 (1998)

\bibitem{carrington}A.~T. Carrington  \textit{et al}., Phys. Rev. B 59, R14173 (1999)

\bibitem{prozorov2}R.~Prozorov, R.W. Giannetta, A. Carrington, F.M. Araujo-Moreira, Phys.
Rev B 62, 115 (2000).

\bibitem{hardy}W.~N. Hardy, D.A. Bonn, D.~C. Morgan, Phys. Rev. Lett. 70, 3999
(1993)

\bibitem{annet}J.~F. Annet, N.D. Goldenfeld and S. Renn, in Physicsl
Properties of High Temperature Superconductors II, edited by D.M. Ginzburg (World Scientific, New Jersey,
1990).

\bibitem{annet2}J.~F. Annet, N. Goldenfeld and A.J. Leggett J. Low Temp. Phys. 105, 473
(1996)

\bibitem{hirschfeld}P.~J. Hirschfeld, W.O. Putikka, D.J. Scalapino, Phys. Rev. B 50, 10250
(1994)

\bibitem{li}Mei-Rong Li, P.~J. Hirschfeld and P. Wolfle Phys. Rev. B 61, 648 (2000)

\bibitem{carrington2}A.~Carrington \textit{et al}., Phys. Rev. Lett., 4172 (1999)

\end{thebibliography}
\end{document}